\documentclass[conference]{IEEEtran}
\IEEEoverridecommandlockouts
\usepackage{cite}
\usepackage{subcaption}
\usepackage{textcomp}
\usepackage{xcolor}
\usepackage{bm}
\usepackage{multirow}
\usepackage{times}
\usepackage{caption}
\usepackage{cite}
\usepackage{amsmath,amssymb,amsfonts}
\usepackage{graphicx}
\usepackage{amsthm}

\usepackage{amsfonts}
\usepackage{cite,graphicx,amsmath,amsthm}
\usepackage{fancyhdr}
\usepackage{dsfont}
\usepackage{array,color}
\usepackage{bm}
\usepackage{float}
\usepackage{algpseudocode}
\usepackage{multirow}
\usepackage{booktabs}
\usepackage{makecell}
\usepackage{hyperref}
\usepackage{graphicx}
\usepackage{pifont}

\allowdisplaybreaks[4]

\newtheorem{lemma}{Lemma}

\newtheorem{proposition}{Proposition}

\newtheorem{corollary}{Corollary}

\newtheorem{property}{Property}

\newtheorem{remark}{Remark}

\newtheorem{claim}{Claim}

\usepackage[table]{xcolor}

\usepackage{flushend}

\begin{document}

\title{
\huge Planning Oriented Integrated Sensing and Communication
\vspace{-0.1in}
\author{
Xibin Jin, Guoliang Li, Shuai Wang$^{\dag}$, Fan Liu, Miaowen Wen$^{\dag}$, Huseyin Arslan, \emph{Fellow, IEEE},\\Derrick Wing Kwan Ng, \emph{Fellow, IEEE}, and Chengzhong Xu, \emph{Fellow, IEEE}
\vspace{-0.1in}
\thanks{
Xibin Jin and Miaowen Wen are with the South China University of Technology, Guangdong, China.
Guoliang Li and Chengzhong Xu are with the University of Macau, Macau, China.
Shuai Wang is with the Shenzhen Institutes of Advanced Technology, Chinese Academy of Sciences, Guangdong, China. 
Fan Liu is with Southeast University, Jiangsu, China.
Huseyin Arslan is with Istanbul Medipol University, Istanbul, Turkey.
Derrick Wing Kwan Ng is with the University of New South Wales, Sydney, Australia.

Corresponding authors: Shuai Wang and Miaowen Wen. 
}
}
}

\maketitle

\begin{abstract}
Integrated sensing and communication (ISAC) enables simultaneous localization, environment perception, and data exchange for connected autonomous vehicles. However, most existing ISAC designs prioritize sensing accuracy and communication throughput, treating all targets uniformly and overlooking the impact of critical obstacles on motion efficiency. To overcome this limitation, we propose a planning-oriented ISAC (PISAC) framework that reduces the sensing uncertainty of planning-bottleneck obstacles and expands the safe navigable path for the ego-vehicle, thereby bridging the gap between physical-layer optimization and motion-level planning. The core of PISAC lies in deriving a closed-form safety bound that explicitly links ISAC transmit power to sensing uncertainty, based on the Cramér–Rao Bound and occupancy inflation principles. Using this model, we formulate a bilevel power allocation and motion planning (PAMP) problem, where the inner layer optimizes the ISAC beam power distribution and the outer layer computes a collision-free trajectory under uncertainty-aware safety constraints. Comprehensive simulations in high-fidelity urban driving environments demonstrate that PISAC achieves up to $40\%$ higher success rates and over $5\%$ shorter traversal times than existing ISAC-based and communication-oriented benchmarks, validating its effectiveness in enhancing both safety and efficiency.

\end{abstract}


\vspace{-0.15in}
\section{Introduction}

Vehicle motion planning aims to generate goal-oriented trajectories while satisfying safety constraints \cite{Neupan}. However, standalone planning systems often degrade in complex or adversarial environments due to sensing uncertainties, where local perception data may be incomplete or inaccurate \cite{9561612,pei2023collaborative}. 
Leveraging complementary measurements from roadside units (RSUs) can effectively mitigate these uncertainties and enhance motion reliability \cite{EdgeV2X}, thereby motivating extensive research on connected autonomous vehicles (CAVs) \cite{9793623}.

To enhance both spectral and energy efficiency in CAVs, the paradigm of integrated sensing and communication (ISAC) has recently emerged \cite{RdarSignalProcess,Radar_Centric, RadarEstimate,Meng2023VehicularConnectivity,Dou2024IntegratedSensing}. By utilizing the same signals for both radar sensing and data communication, ISAC eliminates the traditional separation between target localization and information transmission at the physical layer.
Existing ISAC methods often adopt sensing accuracy \cite{Meng2023VehicularConnectivity}, communication rate \cite{zhang2024efficient}, user fairness \cite{Dou2024IntegratedSensing}, or their combinations \cite{Radar_Centric}, as design objectives. 
However, such metrics are not tailored to end-to-end motion planning and fail to capture the topological dependencies between obstacle vehicles (OVs) and the ego-vehicle (EV). 
As such, the methods \cite{Radar_Centric, zhang2024efficient, RadarEstimate,Meng2023VehicularConnectivity,Dou2024IntegratedSensing} exhibit degraded planning performance in cluttered or obstacle-dense environments.

\begin{figure}[!t]
\centering 
\includegraphics[width=0.95\linewidth]{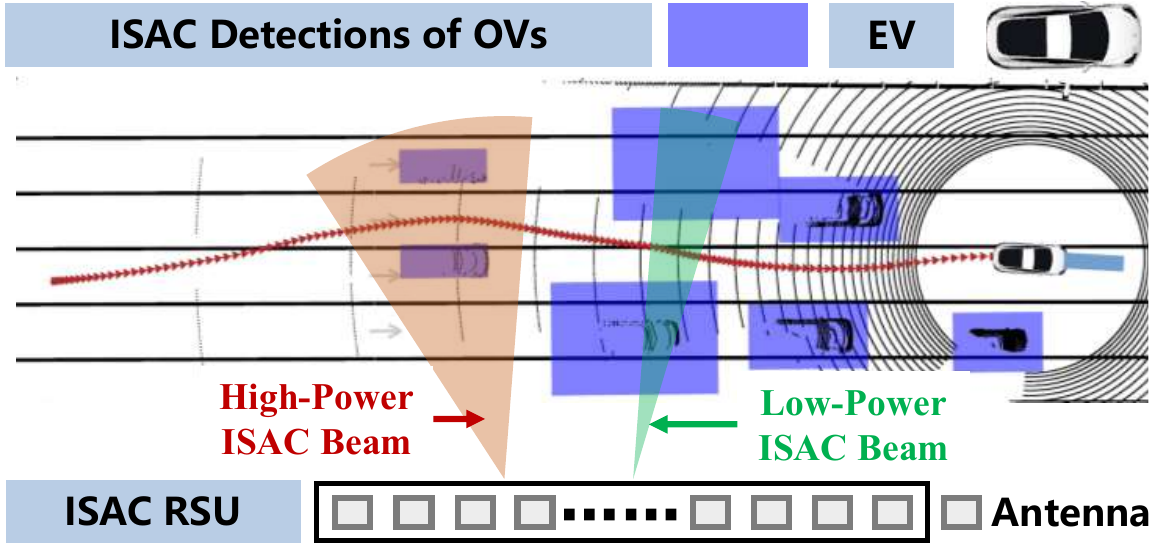}
\caption{The CAV system with $1$\,EV, $K$\,OVs, and $1$\,ISAC RSU.}
\label{fig1}
\vspace{-0.15in}
\end{figure}

To fill this gap, we propose a planning-oriented ISAC (PISAC) framework, which maximizes the planning performance under resource constraints, thereby transcending conventional ISAC designs. 
Specifically, we first derive a closed-form safety bound as a function of ISAC transmit power, leveraging the Cramér-Rao Bound (CRB) \cite{RdarSignalProcess} and occupancy inflation.
We introduce a safety-space shrinkage thresholding model \cite{schulman2014motion}, since the planning performance (i.e., time efficiency and success rate) increases with the safety space \cite{Neupan,schulman2014motion}.
By minimizing this model, PISAC prioritizes the reduction of sensing uncertainty for planning-bottleneck obstacles, in contrast to conventional ISAC approaches \cite{Radar_Centric, zhang2024efficient, RadarEstimate,Meng2023VehicularConnectivity,Dou2024IntegratedSensing} that treat different targets equally.
In doing so, PISAC effectively “illuminates” the driving corridor for the ego-vehicle, as illustrated in Fig. \ref{fig1}. 

Subsequently, we integrate the newly derived objective function into a model predictive control framework to enable cross-layer optimization within PISAC. This leads to a bilevel power allocation and motion planning (PAMP) framework, which unifies sensing, communication, and planning into a single optimization.
In particular, for the inner-level PA, we exploit double circle approximation (DCA) \cite{TwoStagePlanner} and determines the powers of ISAC beams using convex optimization \cite{CVXPY}. 
For the outer-level MP, we compute the collision-free trajectory under ISAC uncertainties using alternating direction method of multipliers (ADMM) \cite{RDA}.

We evaluate the proposed framework in the high-fidelity CARLA (Car Learning to Act) simulator \cite{dosovitskiy2017carla} implemented in Python. 
Experimental results in urban driving scenarios confirm the superior performance of PISAC over a wide range of benchmark schemes. In particular, PISAC reduces traversal time by more than $5\%$ and improves the success rate by $40\%$ compared to conventional ISAC optimization \cite{Radar_Centric, Meng2023VehicularConnectivity}, sum-rate maximization (SRM) \cite{zhang2024efficient}, and max-min fairness (MMF) schemes \cite{Dou2024IntegratedSensing}. 

\vspace{-0.05in}
\section{System Model}\label{section2}
\vspace{-0.03in}
We consider a CAV system consisting of $1$ EV, $K$ OVs, and $1$ ISAC RSU with $N_t$ transmit antennas and $N_r$ receive antennas ($N_t, N_r\gg 1$), as shown in Fig.~\ref{fig1}. 
The RSU adopts a uniform linear array\footnote{Without loss of generality, we assume that the OVs' orientations are all parallel to the antenna array of the RSU.} with half-wavelength spacing, and sends downlink ISAC signals $\tilde{\mathbf{s}}\!=\![\tilde{s}_{1}(t),\dots,\tilde{s}_{K}(t)]^{T}\!\in\!\mathbb{C}^{K}$ to simultaneously provide position estimation and communication services for the OVs in the set $\mathcal{K}=\{1,\dots,K\}$. 
The EV then plans its trajectory based on the ISAC-detected positions of the OVs in $\mathcal{K}$.

\vspace{-0.06in}
\subsection{ISAC Signal Model}
The transmitted signal of the RSU is $\overline{\mathbf{s}}=\mathbf{F}\tilde{\mathbf{s}} \in \mathbb{C}^{N_{t}}$, where $\mathbf{F}\!=\![ \mathbf{f}_{1},\mathbf{f}_{2},\dots,\mathbf{f}_{K}]\!\in\! \mathbb{C}^{N_t \times K}$ denotes the RSU's transmit beamforming matrix. 
The RSU forms the ISAC beam for the $k$-th OV in $\mathcal{K}$ by adopting the transmit beamformer $\mathbf{f}_{k}=\mathbf{a}(\theta_k)\!\in\!\mathbb{C}^{N_t}$ and receive beamformer $\mathbf{r}_{k}=\mathbf{b}(\theta_k)\!\in \!\mathbb{C}^{N_r}$, where $\theta_{k}$ denotes the azimuth angle of the $k$-th OV relative to RSU, and $(\mathbf{a},\mathbf{b})$ are steering vectors.
The radar echo signal $\mathbf{y}_{k}^{R}(t)$ at RSU and the received communication signal $y_{k}^{C}(t)$ at OV are given by\footnote{We assume negligible inter-beam interference owing to the asymptotic orthogonality property of massive antenna arrays \cite{Tensor_mMIMO}.}
\begin{subequations}
\begin{align}
\mathbf{y}_{k}^{R}(t) &\!=\!\kappa_R\beta_{k}\sqrt{p_{k}}\mathbf{b}(\theta_{k})\mathbf{a}^{H}(\theta_{k})\mathbf{f}_{k}\tilde{s}_{k}(t\!-\!\tau_{k})\!+\!\mathbf{z}_{k}^{R}(t), \label{yr}
            \\
y_{k}^{C}(t) &\!=\!\kappa_C\alpha_{k}\sqrt{p_{k}}\mathbf{a}^{H}(\theta_{k})\mathbf{f}_{k}\tilde{s}_{k}(t-\tau_{k}/2)+z_{k}^{C}(t),  \label{yc}
\end{align}
\end{subequations}
where $\kappa_R\!=\!\sqrt{N_{t}N_{r}}$ and $\kappa_C=\sqrt{N_{t}}$ denote the array gains, $\beta_{k}$ is the reflection coefficient,  
$\alpha_{k}$ is the channel coefficient, 
$p_{k}$ is the transmit power of the $k$-th beam, $\tau_{k}$ denotes the round-trip delay, and the symbols $\mathbf{z}_{k}^{R}(t) \in \mathbb{C}^{N_r}$ and $z_{k}^{C}(t)$ denote additive white Gaussian noise (AWGN) with variances of $\sigma_{R}^{2}$ and $\sigma^{2}_{C}$.

Assuming line-of-sight (LoS) transmission, the channel coefficient is given by $\alpha_{k}=\alpha_{\text{ref}}d^{-1}_{k}e^{j\frac{2\pi f_c}{c}d_{k}}$, where $d_{k}$ denotes the distance between $k$-th OV and RSU, $f_{c}$ and $\alpha_{\text{ref}}$ denote the carrier frequency and the known reference path-loss at the unit distance, respectively.
Moreover, $\beta_{k}$ and $\tau_{k}$ can be calculated as $\beta_{k}=\xi/2d_{k}$ \cite{RdarSignalProcess} and $\tau_k=2d_{k} /c$, where $\xi$ is the complex radar cross section and $c$ is the speed of the light.
Lastly, to simplify the notation, we define beam gain $\delta_{k} = \mathbf{a}^{H}(\theta_{k})\mathbf{f}_{k}$.

\vspace{-0.07in}
\subsection{Motion Planning Model}
\vspace{-0.03in}
We adopt the MPC framework for motion planning. 
The operation of MPC is divided into $H+1$ time slots, constituting $\mathcal{H} = \{0, \ldots, H\}$, and the period between consecutive slots is $\Delta T$. 
At the $t$-th time slot ($t \in \mathcal{H}$), the EV's state is denoted as $\mathbf{s}_{t}^{E} = (x_t^E, y_t^E, \theta_t^E)$, where $(x_t^E, y_t^E)$ and $\theta_t^E$ are position and orientation, respectively. 
The OVs' states are $\mathbf{s}_{k}$ without index $E$.
The motion of the EV follows the state evolution model $\mathbf{s}_{t+1}^E = f(\mathbf{s}_t^E, \mathbf{u}_t^E)$, where $\mathbf{u}_t^E$ is the control action of EV and 
\begin{equation}
    f(\mathbf{s}_t^E, \mathbf{u}_t^E) = \mathbf{A}_t \mathbf{s}_t^E + \mathbf{B}_t \mathbf{u}_t^E + \mathbf{c}_t, \quad \forall t \in \mathcal{H}, \label{dynamics}
\end{equation}
with coefficient matrices $(\mathbf{A}_t, \mathbf{B}_t, \mathbf{c}_t)$ determined by Ackermann kinetics \cite[Eq. (8)-(10)]{RDA}. The control vector is bounded by $\mathbf{u}_{\text{min}} \!\preceq\!\mathbf{u}_t^E\!\preceq\!\mathbf{u}_{\text{max}},\forall t$, and $\mathbf{a}_{\text{min}} \preceq \mathbf{u}_{t+1}^E - \mathbf{u}_t^E \preceq \mathbf{a}_{\text{max}}, \forall t$, where $\mathbf{u}_{\text{min}}$ and $\mathbf{u}_{\text{max}}$ are the minimum and maximum values of the control vector, respectively, and $\mathbf{a}_{\text{min}}$ and $\mathbf{a}_{\text{max}}$ are the associated minimum and maximum accelerations, respectively.

Given the initial vehicle state $\mathbf{s}_{0}^{E}=\mathbf{s}_{\mathrm{start}}$, we aim to generate a sequence of EV states $\{\mathbf{s}_{t}^{E}\}_{t\in\mathcal{H}}$ such that 
$\mathbf{s}_{H}^{E}=\mathbf{s}_{\mathrm{goal}}$, where 
$\mathbf{s}_{\mathrm{goal}}$ is the goal position provided by human commands \cite{Neupan}. 
Accordingly, the cost function to be minimized is
\begin{align}
C_0(\{\mathbf{s}_t^E, \mathbf{u}_t^E\}_{t=0}^{H}) = \sum_{t=0}^{H} ||\mathbf{s}_{t}^E-\mathbf{s}_{t}^{\diamond}||_{2}^{2},\label{MPC}
\end{align}
where the discrete waypoints $\{\mathbf{s}_0^{\diamond}, \ldots, \mathbf{s}_{H}^{\diamond}\}$ are sampled from the line $\mathbf{s}_{\mathrm{start}}\rightarrow\mathbf{s}_{\mathrm{goal}}$ based on a given vehicle speed \cite{Neupan}.

\vspace{-0.08in}
\section{PISAC Problem Formulation}
\vspace{-0.05in}
In the considered CAV system, we must guarantee no collision between EV and OVs.
Let $\mathbb{G}_{t}$ and $\mathbb{O}_{k}$ denote the regions occupied by the EV and the $k$-th OV, respectively.
Their set representations are written as \cite{Neupan}
\begin{subequations}
\begin{align}
\mathbb{G}_t(\mathbf{s}^{E}_t) &= \{\mathbf{z} \in \mathbb{Z} | \mathbf{R}(\mathbf{s}_t^{E})\mathbf{z} + \mathbf{p}(\mathbf{s}_t^{E})\}, \forall t  \\
\mathbb{Z} &= \{\mathbf{z} \in \mathbb{R}^{2} | \mathbf{G}\mathbf{z} \leq \mathbf{g}\}, \nonumber\\
\mathbb{O}_{k}(\mathbf{s}_{k}, \mathbf{l}_{k}) &= \{\mathbf{o} \in \mathbb{R}^{2} | \mathbf{D}_{k}(\mathbf{s}_{k},\mathbf{l}_{k})\mathbf{o} \leq \mathbf{b}_{k}(\mathbf{s}_{k},\mathbf{l}_{k}) \}, \forall k,
\end{align}
\end{subequations}
where $\mathbf{R}(\mathbf{s}_t^E) \in \mathbb{R}^{2 \times 2}$ is the rotation matrix related to $\theta_t^E$ of EV and $\mathbf{p}(\mathbf{s}_t^E) \in \mathbb{R}^{2 \times 1}$ is the translation vector related to $(x_t^E, y_t^E)$. The set $\mathbb{Z}$ represents the initial EV shape, where $[\mathbf{G}, \mathbf{D}_{k}] \in \mathbb{R}^{4 \times 2}$, $[\mathbf{g}, \mathbf{b}_{k}] \in \mathbb{R}^4$ (i.e., both EV and OVs are modeled as rectangular rigid bodies), and $\mathbf{l}_{k}=[L_k^1,L_k^2]^{T}$ denotes the length and width of the $k$-th OV, respectively. 

With the above model, the distance between $\mathbb{G}_{t}(\mathbf{s}_{t}^E)$ and $\mathbb{O}_{k}({\mathbf{s}}_{k},\mathbf{l}_{k})$ can be written as
\begin{align}
&\text{dis}(\mathbb{G}_{t}(\mathbf{s}_{t}^E),\mathbb{O}_{k}({\mathbf{s}}_{k},\mathbf{l}_{k})) 
\nonumber\\
&=\min \{\|\mathbf{e}\|_2\!\mid\!(\mathbb{G}_t(\mathbf{s}_{t}^E)\!+\!\mathbf{e}) \cap  \mathbb{O}_{k}({\mathbf{s}}_{k},\mathbf{l}_{k}) \!\neq \!\emptyset\},
\end{align}
which should be larger than a safety margin $d_{\text{safe}}$ during the planning horizon $\mathcal{H}$ \cite{Neupan}.
However, since $\mathbf{s}_{k}$ is estimated from ISAC signal $\mathbf{y}_{k}^{R}(t)$, there is an inevitable error between the estimated $\hat{\mathbf{s}}_{k}$ and the actual $\mathbf{s}_{k}$, which is the so-called ISAC uncertainties and results in $\mathbb{O}_{k}(\hat{\mathbf{s}}_{k},\mathbf{l}_{k}) \neq \mathbb{O}_{k}(\mathbf{s}_{k},\mathbf{l}_{k})$.

Under ISAC uncertainty, the OV state $\mathbf{s}_k$  can be modeled as a random variable following 
\begin{align}
\mathbf{s}_{k} \sim \mathcal{N}(\hat{\mathbf{s}}_{k},\mathbf{\Sigma_{\mathbf{s}_{k}}}), 
\ 
\mathbf{\Sigma_{\mathbf{s}_{k}}} \!=\! \mathbb{E}[(\mathbf{s}_{k}-\hat{\mathbf{s}}_{k})(\mathbf{s}_{k}-\hat{\mathbf{s}}_{k})^{T}], \label{distribution}
\end{align}
where the mean $\hat{\mathbf{s}}_{k}$ and covariance matrix $\mathbf{\Sigma_{\mathbf{s}_{k}}}$ are related to the ISAC beam power $p_k$ as seen from \eqref{yr}. 
By adopting chance-constrained optimization \cite{ChanceConstraint}, the collision avoidance constraint can be expressed probabilistically as
\begin{align}
&\text{Pr}(\text{dis}(\mathbb{G}_t(\mathbf{s}_t^E), \mathbb{O}_{k}(\mathbf{s}_{k},\mathbf{l}_{k}))\!\leq\!d_{\text{safe}}| p_k)
\leq p_{\epsilon}, \ \forall t, k, \label{chance}
\end{align}
where $p_{\epsilon}$ is a small risk threshold.

Having the planning constraints satisfied, it is then crucial to ensure the communication performance at each OV:
\begin{align}
\sum_{k=1}^{K} \log_{2}\left(1 + p_{k}\kappa^{2}_C|\alpha_{k}|^{2}|\delta_{k}|^{2}/\sigma^{2}_{C}\right) \geq R_0, \label{rate}
\end{align}
where $R_0$ denotes the required downlink sum-rate threshold. 

The PISAC problem is thus formulated as 
\begin{subequations}
\begin{align}
\mathsf{P}_0:~&\min_{\{\mathbf{s}_{t}^E, \mathbf{u}_{t}^E\}_{t=0}^{H}, \{p_k\}_{k=0}^K}~~ \sum_{t=0}^{H} ||\mathbf{s}_{t}^E-\mathbf{s}_{t}^{\diamond}||_{2}^{2} \label{P0_1}   \\
\text{s.t.}~~&\mathbf{s}_{t+1}^E = \mathbf{A}_t \mathbf{s}_t^E + \mathbf{B}_t \mathbf{u}_t^E + \mathbf{c}_t, \quad \forall t \in \mathcal{H}, \label{P0_2}  \\
&\mathbf{u}_{\text{min}} \leq \mathbf{u}_t^E \leq \mathbf{u}_{\text{max}}, \quad \forall t \in \mathcal{H},  \label{P0_3}\\
&\mathbf{a}_{\text{min}} \leq \mathbf{u}_{t+1}^E - \mathbf{u}_t^E \leq \mathbf{a}_{\text{max}}, \forall t \in \mathcal{H}, \label{P0_4} 
\\
&p_{k} \geq 0, \ \sum_{k=1}^{K}p_{k}\leq P_{\text{sum}}, \label{P0_5}
\\
& \mathrm{constraints}~\eqref{chance}, \eqref{rate},
\end{align}
\end{subequations}
where $P_{\text{sum}}$ denotes the total transmit power budget.

\addtolength{\topmargin}{0.01in}
Problem $\mathsf{P}_0$ is inherently non-trivial due to the strong interdependence between power allocation $\{p_k\}_{k=1}^K$ and motion planning $\{\mathbf{s}_{t}^E, \mathbf{u}_{t}^E\}_{t=0}^{H}$. 
This interdependence involves integral and cannot be explicitly calculated as seen from \eqref{chance}.
Existing CAV solutions \cite{Neupan,RDA} simplifies \eqref{chance} into a deterministic form without ISAC uncertainties, while conventional ISAC methods \cite{Radar_Centric, zhang2024efficient, RadarEstimate,Meng2023VehicularConnectivity,Dou2024IntegratedSensing} optimize $\{p_k\}_{k=1}^K$ by minimizing sensing errors or maximizing communication rate, instead of minimizing planning cost \eqref{P0_1}.
All these methods will result in degraded planning performance in obstacle-rich environments.
In contrast, we will solve $\mathsf{P}_0$ directly, by converting it into a tractable form while ensuring its planning oriented nature.

\section{Safety Bound Under ISAC Uncertainties}\label{section3}

In this section, we derive a safety bound to quantify the relation between the chance constraint \eqref{chance} and the ISAC power. 
We first derive the CRB to quantify the ISAC estimation errors.
We then bound the probability in \eqref{chance} by obstacle inflation.

\subsection{Cramér-Rao Bound}
\label{sec:PE}
To tackle the chance constraint \eqref{chance}, we first need to derive the covariance matrix $\mathbf{\Sigma_{\mathbf{s}_{k}}}$ for random variable $\mathbf{s}_k$.
The RSU estimates $\mathbf{s}_k$ from $\mathbf{w}_{k}=[\theta_{k},d_{k}]^{T}$, where $\theta_{k}$ and $d_{k}$ denote yaw and distance in the polar coordinate system, respectively, and $\{\mathbf{s}_{k},\mathbf{w}_{k}\}_{k\in\mathcal{K}}$ satisfy
\begin{align}
x_{k}=d_{k}\cos\theta_{k}, 
\
y_{k}=d_{k}\sin\theta_{k}, \quad \forall k \in \mathcal{K}\label{Transformation}.
\end{align}
Specifically, to measure $\mathbf{w}_{k}$ from \eqref{yr}, we adopt the standard radar signal processing in \cite{RdarSignalProcess}
\begin{align}
\begin{gathered}
\mathbf{r}_{k} \triangleq \mathbf{g}(\mathbf{w}_{k})\!+\!\mathbf{z}_{k}
\!=\!
\begin{bmatrix}
\kappa_R\beta_{k}
\mathbf{b}(\theta_{k})\mathbf{a}^{H}(\theta_{k})\mathbf{a}(\theta_{k}) \\ \frac{2d_{k}}{c}  
\end{bmatrix}
\!+\! 
\begin{bmatrix}
\mathbf{z}_{{k}}^{\theta} \\ z_{{k}}^{\tau}
\end{bmatrix},
\end{gathered} \nonumber
\end{align}
where $\mathbf{z}_{k} \sim \mathcal{CN}(\mathbf{0},\mathbf{Q}_{k})$, $\mathbf{Q}_{k}=\text{diag}(\sigma^{2}_{\theta_{k}}\mathbf{1}^{T}_{N_{r}},\sigma^{2}_{\tau_{k}})$, and 
\begin{align}
\sigma^{2}_{\theta_{k}}=p_{k}^{-1}\cdot\frac{a_2^{2}\sigma_{R}^{2}}{G}, \ \sigma^{2}_{\tau_{k}}=p_{k}^{-1}\cdot\frac{a_1^{2}\sigma_{R}^{2}}{G\kappa_{R}^{2}|\beta_{k}|^2\delta_{k}^2}.
\end{align}
Here, $a_1$, $a_2$ are constant scaling factors, and $G$ is the matched filtering gain.

We exploit the first-order Taylor expansion of $\mathbf{g}(\mathbf{w}_{k})$ at a certain point $\mathbf{w}_{k}^{\text{pre}}$ to eliminate the nonlinearity in the measurement function $\mathbf{g}(\mathbf{w}_{k})$, i.e., $\mathbf{g}(\mathbf{w}_{k}) \approx \mathbf{g}(\mathbf{w}_{k}^{\text{pre}})+\mathbf{U}_{k}(\mathbf{w}_{k}-\mathbf{w}_{k}^{\text{pre}})$, where $\mathbf{U}_{k}\!=\!\frac{\partial \mathbf{g}}{\partial \mathbf{w}_{k}} |_{\mathbf{w}_{k}=\mathbf{w}_{k}^{\text{pre}}}$. By denoting $\tilde{\mathbf{r}}_{k}=\mathbf{r}_{k}-\mathbf{g}(\mathbf{w}_{k}^{\text{pre}})+\mathbf{U}_{k}\mathbf{w}_{k}^{\text{pre}}$, we have $\tilde{\mathbf{r}}_{k}=\mathbf{U}_{k}\mathbf{w}_{k}+\mathbf{z}_{k}$, which is linear in $\mathbf{w}_{k}$.
Thus, the estimated value of $\mathbf{w}_{k}$ can be acquired through a maximum likelihood estimator \cite{RadarEstimate}, written as 
\begin{align}
\hat{\mathbf{w}}_{k} &= \arg\min_{\mathbf{w}_k} \left( \tilde{\mathbf{r}}_k - \mathbf{U}_k \mathbf{w}_k \right)^H \mathbf{Q}_k^{-1} \left( \tilde{\mathbf{r}}_k - \mathbf{U}_k \mathbf{w}_k \right) \nonumber \\
&=(\mathbf{U}^{H}_{k}\mathbf{Q}^{-1}_{k}\mathbf{U}_{k})^{-1}\mathbf{U}^{H}_{k}\mathbf{Q}^{-1}_{k}
\tilde{\mathbf{r}}_{k},
\end{align}
and $\hat{\mathbf{s}}_{k}$ can be obtained according to \eqref{Transformation}.

By denoting $\mathbf{Q}_{k}=p_{k}^{-1}\tilde{\mathbf{Q}}_{k}$, the Fisher Information Matrix (FIM) related to $\mathbf{w}_{k}$ can be computed as 
\begin{align}
\mathbf{J}_k^{\mathbf{w}} =p_{k}\mathbf{U}_k^H \tilde{\mathbf{Q}}_k^{-1} \mathbf{U}_k \triangleq p_{k}\tilde{\mathbf{J}}_{k}^{\mathbf{w}},
\end{align}
and the FIM related to $\mathbf{s}_{k}$ is derived by the chain rule as $\mathbf{J}_{k}^{\mathbf{s}}=p_{k}\mathbf{A}_{k}\tilde{\mathbf{J}}_{k}^{\mathbf{w}}\mathbf{A}_{k}^{T}$ \cite{ChainRule}, where $\mathbf{A}_{k}$ is a Jacobian matrix in the form of 
\begin{align}
\mathbf{A}_{k}=\!\left[ \frac{-y_{k}}{x_{k}^2+y_{k}^2}, \frac{x_{k}}{x_{k}^2+y_{k}^2}; \frac{x_{k}}{\sqrt{x_{k}^2+y_{k}^2}}, \frac{y_{k}}{\sqrt{x_{k}^2+y_{k}^2}} \right].
\end{align}
Then the CRB matrix of OV $k$ related to $\mathbf{s}_{k}$, which measures the lower bound of the variance of position estimation, can be computated as
\begin{align}
\text{CRB}_{k}&=\mathbf{J}_{k}^{\mathbf{s}^{-1}}=p_{k}^{-1}\cdot\left[\mathbf{A}_{k}\tilde{\mathbf{J}}_{k}^{\mathbf{w}}\mathbf{A}_{k}^{T}\right]^{-1} \nonumber \\&\triangleq p_{k}^{-1}\cdot\mathbf{C}_{k} \preceq \mathbb{E}\left[(\hat{\mathbf{s}}_k - \mathbf{s}_k)(\hat{\mathbf{s}}_k - \mathbf{s}_k)^H\right], \label{CRB_s}
\end{align}
and the CRB of coordinates $x_{k}$ and $y_{k}$ are diagonal entries of $\text{CRB}_{k}$, which are given by
\begin{align}
\mathbb{E}\left(\left(\hat{x}_k - x_k\right)^2\right) \geq p_k^{-1}c_{11}^{k} \triangleq \text{CRB}_{k}^{x}, \nonumber\\
\mathbb{E}\left(\left(\hat{y}_k - y_k\right)^2\right) \geq p_k^{-1}c_{22}^{k} \triangleq \text{CRB}_{k}^{y},\label{CRB}
\end{align}
where $c_{ij}^{k}$ denotes the $(i, j)$-th entry of $\mathbf{C}_{k}$. 

Lastly, assuming that $x_{k}$ and $y_{k}$ are independent of each other, and using \eqref{CRB}, we can approximately obtain
\begin{align}\label{Sigma}
\mathbf{\Sigma_{\mathbf{s}_{k}}} &= \left[{\begin{array}{cc}
     p_k^{-1}c_{11}^{k} & 0 \\
    0& p_k^{-1}c_{22}^{k} \\
\end{array}}\!\right].
\end{align}

\subsection{Probability Bound by Occupancy Inflation}
\label{sec:ProRegion}

Based on the result in \eqref{Sigma}, the left hand side collision probability in \eqref{chance} becomes
\begin{align}
&\text{Pr}(p_k)
\!\triangleq\!\int_{\mathbf{s}_{k}} \mathbb{I}_C(\mathbf{s}_{t}^E, \mathbf{s}_{k}) f(\mathbf{s}_{k}) d\mathbf{s}_{k}, \label{chance2}
\end{align}
where $f(\cdot)$ denotes the probability density function (PDF) for distribution 
$\mathbf{s}_{k} \sim \mathcal{N}(\hat{\mathbf{s}}_{k},\mathbf{\Sigma_{\mathbf{s}_{k}}})$, and $\mathbb{I}_C$ is the indicator function defined as
\begin{align}
\mathbb{I}_C(\mathbf{s}_{t}^E, \mathbf{s}_{k}) = \begin{cases}
1, & \text{if} \  \text{dis}(\mathbb{G}_t(\mathbf{s}_{t}^E), \mathbb{O}_{k}(\mathbf{s}_{k},\mathbf{l}_{k}))<d_{\text{safe}} \\
0, & \text{otherwise}
\end{cases}.
\end{align}
This integration is still nontrival to compute. 
To address this issue, we propose an occupancy inflation method that finds a larger set $\mathbb{O}_{k}^{p}(p_k) \triangleq \mathbb{O}_{k} ({\mathbf{s}}_{k}(p_k),\mathbf{l}_k^p(p_k))$, with $\mathbb{O}_{k}({\mathbf{s}}_{k}, \mathbf{l}_{k}) \subseteq \mathbb{O}_{k}^{p}$ and $\mathbf{l}_{k} \leq \mathbf{l}_k^p$, such that the following deterministic form is a sufficient condition for the primal \eqref{chance} to hold true \cite{TwoStagePlanner}: 
\begin{align}\label{sufficient}
\text{dis}(\mathbb{G}_{t}(\mathbf{s}_{t}^E),\mathbb{O}_{k}^{p}(p_k)) \geq d_{\text{safe}} 
\Rightarrow 
\mathrm{constraint}~\eqref{chance}.
\end{align}

Now, the remaining question is to determine the smallest inflated obstacle region $\mathbb{O}_{k}^{p}(p_k)$ satisfying \eqref{sufficient}. 
To answer this question, we define a set of elliptical regions $\mathbb{B}^p(\boldsymbol{m}, \mathbf{M})$ with $\{\mathbf{M} \in \mathbb{R}^{2 \times 2}$: $\mathbf{M} \succ 0\}$ as
\begin{align}
\mathbb{B}^p(\boldsymbol{m}, \mathbf{M}) \triangleq \{\mathbf{x}\in \mathbb{R}^2 \mid (\mathbf{x} - \boldsymbol{m})^T \mathbf{M}^{-1} (\mathbf{x} \!- \!\boldsymbol{m}) \leq 1\},
\end{align}
and then we have the following property.
\begin{property}
For the Gaussian distributed random variable $\mathbf{s}_{k} \sim \mathcal{N}(\hat{\mathbf{s}}_{k},\mathbf{\Sigma_{\mathbf{s}_{k}}})$, 
$\mathbb{B}^p(\hat{\mathbf{s}}_{k}, \chi_{p_{\epsilon}}^2(2)\mathbf{\Sigma_{\mathbf{s}_{k}}})$ is the ellipsoidal confidence set of probability level $p_{\epsilon}$ for $\mathbf{s}_{k}$ satisfying
\begin{align}
\mathrm{Pr}(\mathbf{s}_{k} \in \mathbb{B}^p(\hat{\mathbf{s}}_{k}, \chi_{p_{\epsilon}}^2(2)\mathbf{\Sigma_{\mathbf{s}_{k}}})) = p_{\epsilon},
\end{align}
where $\chi_p^2(2)$ denotes the p-th quantiles of the chi-square distribution with 2 degrees of freedom.
\end{property}
\begin{proof}
Adopting the standard transformation of $\mathbf{z}=\mathbf{\Sigma}_{\mathbf{s}_{k}}^{-\frac{1}{2}}(\mathbf{s}_{k}-\hat{\mathbf{s}}_{k})$, we have $\mathbf{z} \sim \mathcal{N}(\mathbf{0}_{\tiny2\times1},\mathbf{I}_{\tiny2\times2})$. 
Defining $Y\triangleq\mathbf{z}^{T}\mathbf{z}$, it is clear that $Y \sim \chi^{2}(2)$. Consequently, 
\begin{align}
\text{Pr}(\mathbf{s}_{k} \in \mathbb{B}^p)&\Longleftrightarrow\text{Pr}\underbrace{((\mathbf{s}_{k} - \hat{\mathbf{s}}_{k})^T \mathbf{\Sigma}_{\mathbf{s}_{k}}^{\tiny-1} (\mathbf{s}_{k} - \hat{\mathbf{s}}_{k})}_{\equiv[\mathbf{\Sigma}_{\mathbf{s}_{k}}^{\tiny-\frac{1}{2}}(\mathbf{s}_{k}-\hat{\mathbf{s}}_{k})]^{T}[\mathbf{\Sigma}_{\mathbf{s}_{k}}^{\tiny-\frac{1}{2}}(\mathbf{s}_{k}-\hat{\mathbf{s}}_{k})]}\leq \chi^{2}_{p_{\epsilon}}(2)) \nonumber \\
&\Longleftrightarrow\text{Pr}(\mathbf{z}^{T}\mathbf{z} \leq \chi^{2}_{p_{\epsilon}}(2)) \nonumber \Longleftrightarrow \text{Pr}(Y\leq\chi^{2}_{p_{\epsilon}}(2)).
\end{align}
The property is immediately proved.
\end{proof}

Based on \textbf{Property 1}, we further derive $\mathbb{O}_{k}^{p}$ based on rectangular wrapping, which leads to the following result.

\begin{property}
The smallest $\mathbb{O}_{k}^p$ satisfying \eqref{sufficient} can be tightly wrapped by the rectangle $\mathbb{O}_{k}(\hat{\mathbf{s}}_{k},\mathbf{l}_{k}^p)$ centered on the $\hat{\mathbf{s}}_{k}$ with  $\mathbf{l}_{k}^p=[L_{k}^{1},L_{k}^{2}]$, where $L_{k}^{1}=2(L_k^a+p_{k}^{-1/2}\sqrt{c_{11}^{k}·\chi_{p_{\epsilon}}^2(2)})$ and $L_{k}^{2}=2(L_k^b+p_{k}^{-1/2}\sqrt{c_{22}^{k}·\chi_{p_{\epsilon}}^2(2)})$.
\end{property}

\begin{proof}
This property is based on the Minkowski sum of an ellipse and a rectangle. 
See \cite{MinSum_1} for detailed proof.
\end{proof}

\begin{figure*}[!t]
\centering 
\includegraphics[width=0.98\linewidth]{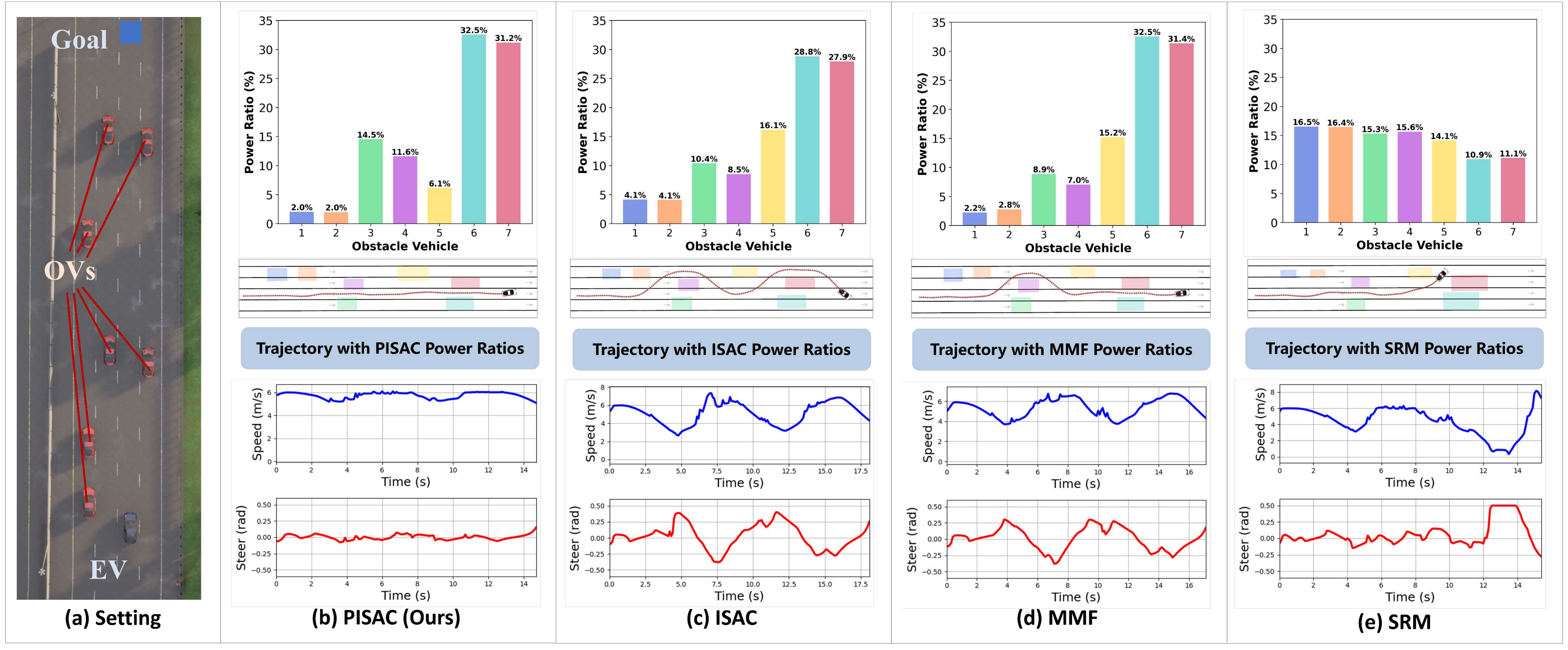}
\vspace{-0.05in}
\caption{Power, trajectory and motion profiles of different schemes, where OVs are indexed by color, and the bbox denotes $\mathbb{O}_k^p$.}
\label{demo}
\vspace{-0.2in}
\end{figure*}

\vspace{-0.06in}
\section{Bilevel PAMP Optimization} \label{section4}

With the results in Sections IV-A and IV-B, we can replace \eqref{chance} by \eqref{sufficient} and solve the resultant optimization problem. 
Specifically, the driving space of EV is defined as the neighborhood area along the EV's intended path \cite[Eq. 11]{li2024edge}:  
\begin{align}
    \mathbb{D}=\{\mathbf{p}: 
    \text{dis}(\mathbb{G}_{t}(\mathbf{s}_{t}^{\diamond}),\mathbf{p}) < d_{\text{safe}} 
    \}.
\end{align}
If an OV lies at $\text{dis}(\mathbb{G}_{t}(\mathbf{s}_{t}^{\diamond}),\mathbb{O}_{k}^{p})<
d_{\text{safe}}
$, the driving space $\mathbb{D}$ would be reduced. 
Otherwise, there is no impact on $\mathbb{D}$.
This results in the safety space shrinkage thresholding cost \cite{schulman2014motion}
\begin{align}
\Xi(\{p_{k}\}_{k=1}^K) = \sum\limits_{t=0}^{H}\sum\limits_{k=0}^{K}\left[d_{\text{safe}}-\text{dis}\left(\mathbb{G}_{t}(\mathbf{s}_{t}^{\diamond}),\mathbb{O}_{k}^{p}(p_k)\right)\right]^{+}, \label{regularized}
\end{align}
where $[x]^{+}=\text{max}(x,0)$.

Based on the above discussions, 
problem $\mathsf{P}_0$ is converted into a bi-level PAMP problem as follows:
\begin{subequations}
\begin{align}
\mathsf{P}_1:~&\min_{\{\mathbf{s}_{t}^E, \mathbf{u}_{t}^E\}_{t=0}^{H}}~~\sum_{t=0}^{H} ||\mathbf{s}_{t}^E-\mathbf{s}_{t}^{\diamond}||_{2}^{2}  \label{P0_11}   \\
\text{s.t.}~~&\{p_{k}^*\}_{\tiny k=1}^{K} = \arg \min_{\{p_k\}_{k=1}^K}\Big\{\Xi(\{p\}_{k=1}^K)+\varphi(\{p_{k}\}_{k=1}^K):\nonumber \\&  
\quad \quad \quad \quad \quad
\mathrm{constraints}~\eqref{rate},\eqref{P0_5}
\Big\}, \label{inner} \\
&\text{dis}(\mathbb{G}_{t}(\mathbf{s}_{t}^E), \mathbb{O}_{k}^{p}(p_k^*)) \geq d_{\text{safe}},\forall t \in \mathcal{H}, \forall k\in\mathcal{K}, \label{P1f} \\
& \mathrm{constraints}~\eqref{P0_2}, \eqref{P0_3}, \eqref{P0_4},
\end{align}
\end{subequations}
where $\varphi(\{p_{k}\}_{k=1}^K)$ is a regularization cost used to ensure the overall ISAC performance \cite {Radar_Centric}:
\begin{align}
\varphi(\{p_{k}\}_{k=1}^{K})=\rho\sum_{k=1}^{K} p_k^{-1} (c_{11}^{k}+c_{22}^{k}), \label{CRB_Min}
\end{align}
and $\rho$ is a hyperparameter to balance the planning and sensing performances. 
Problem $\mathsf{P}_1$ no longer involves the implicit probability constraints. The next subsections present algorithms for solving the inner- and outer-level problems of $\mathsf{P}_1$.

\subsection{Power Allocation via DCA}

To solve the inner PA problem in \eqref{inner}, the key is to handle the function $\text{dis}$ in $\Xi$.
We exploit the DCA method in \cite{TwoStagePlanner} to approximate $\text{dis}(\mathbb{G}_{t}(\mathbf{s}_{t}^{\diamond}),\mathbb{O}_{k}^{p}))$ into a computationally efficient surrogate function. 

Specifically, the
rectangular regions of $\mathbb{G}_{t}(\mathbf{s}_{t}^{\diamond})$ and $\mathbb{O}_{k}^{p}$ are approximated by a set of two identical discs, expressed as
\begin{align}
\mathbb{O}_{k}^{p} \subseteq  \bigcup_{j \in \{1, 2\}} \mathcal{R}_{k,j}^O , \forall k, \  \mathbb{G}_{t}(\mathbf{s}_{t}^{\diamond}) \subseteq \bigcup_{j \in \{1, 2\}} \mathcal{R}_{t,j}^E, \forall t.
\end{align}
We denote the center point of the $j$-th disc of the OV $k$ and EV at state $\mathbf{s}_{t}^{\diamond}$ as $\mathbf{c}_{k,j}^O$ and $\mathbf{c}_{t,j}^E$, and the corresponding radius of the discs are calculated as
\begin{align}
r_{k,j}=r_{k}^0+p_{k}^{-1/2}\sqrt{\chi_{p_{\epsilon}}^{2}(2)(c_{11}^{k}+c_{22}^{k})}, \quad r_{t,j}=r_{0},
\end{align}
where $r_{k}^0$ and $r_{0}$ denote the initial radius of the circle which circumscribed the equally divided rectangle area of OV $k$ and EV, respectively. 

Hence, $\text{dis}(\mathbb{G}_{t}(\mathbf{s}_{t}^{\diamond}),\mathbb{O}_{k}^{p}))$ can be approximately calculated by the minimum distance between two pairs of disc sets, i.e., 
\begin{align}
\text{dis}(\mathbb{G}_{t}(\mathbf{s}_{t}^{\diamond}),\mathbb{O}_{k}^{p}))\!\approx \!\underbrace{\min_{i,j \in \{1,2\}}\!\left\{\|\mathbf{c}_{t,j}^E\!-\!\mathbf{c}_{k,i}^O \|_2\!-\!r_{k,j}\!-\!r_{t,j}\right\}}_{\triangleq \Gamma_{k,t}(p_{k})}.
\end{align}
Accordingly, we have the following
\begin{align}
    \Xi(\{p_k\}) \approx 
    \tilde{\Xi}(\{p_k\})=
    \sum_{t=1}^H\sum_{k=1}^K
\left[d_{\text{safe}}\!-\!\Gamma_{k,t}(p_{k})\right]^{+}.
 \label{drivingspace}
\end{align}
Thus, the inner PA problem \eqref{inner} is approximated into 
\begin{equation}
\begin{aligned}
	\mathsf{P}_{\mathrm{PA}}: \mathop{\mathrm{min}}_{\{p_{k}\}_{k=1}^{K}}~
	&\tilde{\Xi}(\{p_k\}_{k=1}^K)+\varphi(\{p_{k}\}_{k=1}^K) \\
	\mathrm{s.t.}~~~&\mathrm{constraints}~\eqref{rate},\eqref{P0_5}, 
\label{balance}
\end{aligned}
\end{equation}
which is convex and can be efficiently solved by off-the-shelf software packages (i.e, cvxpy) for convex programming with a complexity of $\mathcal{O}(K^{3.5})$.

\subsection{Motion Planning via ADMM}

After acquiring the optimal $\{p_{k}^*\}_{k=1}^{K}$ by solving the problem $\mathsf{P}_{\mathrm{PA}}$, we put $\{p_{k}^*\}_{k=1}^{K}$ into \eqref{P1f} and problem  $\mathsf{P}_{1}$ becomes
\begin{subequations}
\begin{align}
\mathsf{P}_{\mathrm{MP}}:~&\min_{\{\mathbf{s}_{t}^E, \mathbf{u}_{t}^E\}_{t=0}^{H}}~~\sum_{t=0}^{H} ||\mathbf{s}_{t}^E-\mathbf{s}_{t}^{\diamond}||_{2}^{2}  \label{MPa}  \\
\text{s.t.}~~
& \mathrm{constraints}~\eqref{P0_2}, \eqref{P0_3}, \eqref{P0_4}, \\
&\text{dis}(\mathbb{G}_{t}(\mathbf{s}_{t}^E), \mathbb{O}_{k}^{p*}) \geq d_{\text{safe}},\forall t \in \mathcal{H}, \forall k\in\mathcal{K}.  \label{MPe}
\end{align}
\end{subequations}
To solve this problem, we adopt the strong duality method to transform constraint \eqref{MPe} into a more computationally efficient form \cite{OBCA}, i.e., 
\begin{subequations}
\begin{align}
&\text{dis}(\mathbb{G}_{t}(\mathbf{s}_{t}^E),\mathbb{O}_{k}^{p}) \geq d_{\text{safe}} \nonumber\\&\Longleftrightarrow 
\boldsymbol{\lambda}_{k,t} \geq 0, \, \boldsymbol{\mu}_{k,t} \geq 0, \label{dual1} \\ 
&\ \ \ \ \ \ \ \boldsymbol{\lambda}_{k,t}^{\top} \mathbf{D}_{k} \mathbf{p}(\mathbf{s}_t^E) - \boldsymbol{\lambda}_{k,t}^{\top} \mathbf{b}_{k} - \boldsymbol{\mu}_{k,t}^{\top} \mathbf{g} \geq d_{\text{safe}}, \label{dual2} \\
&\phantom{\Longleftrightarrow} \ \ \boldsymbol{\mu}_{k,t}^{\top} \mathbf{G} + \boldsymbol{\lambda}_{k,t}^{\top} \mathbf{D}_{k} \mathbf{R}(\mathbf{s}_t^E) = 0, \, \|\mathbf{D}_{k,t}^{\top} \boldsymbol{\lambda}_{k,t}\|_* \leq 1, \label{dual3}
\end{align}
\end{subequations}
where $\{\boldsymbol{\lambda}_{k,t},\boldsymbol{\mu}_{k,t}\}$ are dual variables. By plugging \eqref{dual1}--\eqref{dual3} into \eqref{MPe}, problem $\mathsf{P}_{\mathrm{MP}}$ becomes 
\begin{subequations}
\begin{align}
\mathsf{P}_{\mathrm{MP}}':~&\min_{\{\mathbf{s}_{t}^E, \mathbf{u}_{t}^E\},
\{\boldsymbol{\lambda}_{k,t},\boldsymbol{\mu}_{k,t}\}
}~~\sum_{t=0}^{H} ||\mathbf{s}_{t}^E-\mathbf{s}_{t}^{\diamond}||_{2}^{2}  \\
\text{s.t.}~~&\mathrm{constraints}~\eqref{P0_2}, \eqref{P0_3}, \eqref{P0_4},\eqref{dual1},\eqref{dual2},\eqref{dual3}.
\end{align}
\end{subequations}
This is a bi-convex optimization problem with respect to $\{\mathbf{s}_t^E,\mathbf{u}_t^E,\boldsymbol{\lambda}_{k,t},\boldsymbol{\mu}_{k,t}\}$ \cite{RDA}, i.e., given fixed $\{\mathbf{s}_t^E,\mathbf{u}_t^E\}$, the subproblem of $\{\boldsymbol{\lambda}_{k,t},\boldsymbol{\mu}_{k,t}\}$ is convex and vice versa. 
According to \cite{RDA}, we can exploit ADMM to solve $\mathsf{P}_{\mathrm{MP}}'$, which iterates between solving $\{\mathbf{s}_t^E,\mathbf{u}_t^E\}$'s subproblem and $\{\boldsymbol{\lambda}_{k,t},\boldsymbol{\mu}_{k,t}\}$'s subproblem. 
Each subproblem can be solved by cvxpy \cite{CVXPY}. 
The ADMM procedure converges to a stationary point of $\mathsf{P}_{\mathrm{MP}}'$ (thus $\mathsf{P}_{\mathrm{MP}}$) with a complexity of $\mathcal{O}(\mathcal{I}\left[(5H)^{3.5}+HK(8)^{3.5}\right])$ and $\mathcal{I}$ is the number of iterations for ADMM to converge.

\section{Simulation Results}\label{section5}

\begin{table}[!t]
\centering
\caption{Planning performances at different SNRs.}
\vspace{-0.08in}
\scalebox{0.72}{
\begin{tabular}{ccccccc}
\toprule
\multirow{2}{*}{\textbf{SNR}} & \multirow{2}{*}{\textbf{Method}} & \multicolumn{1}{c}{\textbf{AvgAcc}\,$\downarrow$} & \multicolumn{1}{c}{\textbf{MaxAcc}\,$\downarrow$} & \multicolumn{1}{c}{\textbf{PassTime}\,$\downarrow$} & \multicolumn{1}{c}{\textbf{TrajLength}\,$\downarrow$} & \multicolumn{1}{c}{\textbf{SuccessRate}\,$\uparrow$} \\
 & & (m/s\(^2\)) & (m/s\(^2\)) & (s) & (m) & (\%) \\
\midrule
\multirow{3}{*}{$36\,$dB} 
 & ISAC [7] & 1.071 & 8.503  & 17.220  & 90.386 & 60.0 \ (12/20) \\
 & SRM [11] & 1.056 & 7.763  & 16.510 & 88.802 & 50.0 \ (10/20) \\
 & MMF [10] & 1.189 & 7.915  & 16.850 & 87.995 & 55.0 \ (11/20) \\
 \rowcolor{blue!10} & PISAC(Ours) & \textbf{0.994} & \textbf{7.408} & \textbf{15.790} & \textbf{84.564} & \textbf{100.0 (20/20)} \\
\midrule
\multirow{3}{*}{$38\,$dB} 
 & ISAC [7] & 0.994 & 7.481  & 16.710 & 90.653 & 100.0 \ (20/20) \\
 & SRM [11] & 1.034 & 6.712  & 16.220  & 87.532 & 65.0 \ (13/20) \\
 & MMF [10] & 1.117 & 7.850  & 16.840 & 89.182 & 75.0 \ (15/20) \\
 \rowcolor{blue!10} & PISAC(Ours) & \textbf{0.833} & \textbf{6.601} & \textbf{15.530} & \textbf{84.573}  & \textbf{100.0 (20/20)} \\ 
\bottomrule
\end{tabular}
}
\label{perform}
\vspace{-0.25in}
\end{table}

We implemented PISAC using Python in unreal engine Carla simulator \cite{dosovitskiy2017carla} on a Linux workstation with 3.7\,GHz AMD CPU and NVIDIA 3090\,Ti GPU.
As shown in Fig.~\ref{demo}a, our experiments are conducted in a multilane urban driving scenario in Town04 of Carla with $K=7$. 
The positions $\mathbf{s}_{\text{start}}$ and $\mathbf{s}_{\text{goal}}$ are set to $(409.2, 28)$ and $(409.2, 113)$, respectively. 
The safety distance is set to $d_{\text{safe}}=0.15$\,m. The length of planning horizon is $H=20$, with a time step of $\Delta T = 0.1$\,s. All vehicles are Tesla Model3 with length $L_k^a=4.694$m and width $L_k^b=1.849$m. 
The ISAC RSU is located at $(380, 38.5)$, and its number of antennas is $N_t=N_r=64$. 
We define the transmit SNR as $P_\text{sum}/\sigma_R^2$ \cite{RdarSignalProcess}, with unit noise variances $\sigma_R^2=\sigma_C^2=1$. 
We set the matched filtering gain $G=10$, $a_1=6.7\times10^{-5}$, $a_2=1$, and $\xi=1+1j$.

We compare our PISAC against the following baselines: 
1) \textbf{ISAC}: ISAC power allocation based on CRB minimization \cite{Radar_Centric}; 
2) \textbf{SRM}: Power allocation based on sum-rate maximization \cite{zhang2024efficient}; 
3) \textbf{MMF}: Power allocation based on max-min fairness \cite{Dou2024IntegratedSensing}; 
4) \textbf{RDA}: A sota MPC planner without consideration of ISAC uncertainties \cite{RDA}. 

First, the power allocation, trajectory, and motion profiles of all schemes are illustrated in Fig.~\ref{demo}b–Fig.~\ref{demo}e. 
It can be seen from Fig.~\ref{demo}b that PISAC finds the shortest path to reach the goal within $15\,$s and the steering angle maintains small fluctuations. This is realized by prioritizing ISAC powers to planning bottleneck OVs $[3,4,6,7]$ as shown in upper part of Fig.~\ref{demo}b. As such, their estimation uncertainties are reduced, ensuring enough room for EV to drive through the gap. 
In contrast, the conventional ISAC scheme minimizes only the average sensing uncertainty, neglecting the spatial importance of individual obstacles, while the MMF approach allocates excessive power to distant OVs.
These baselines result in longer trajectories and larger steering, as seen from Fig.~\ref{demo}c–Fig.~\ref{demo}d. 
Lastly, SRM exhibits water-filling power profiles, and fails the task with stuck, as shown in Fig.~\ref{demo}e.

Next, quantitative results are summarized in Table I, where each data point represents the average of 20 independent simulation runs under different noises. 
The proposed PISAC achieves the highest success rate (i.e., $100\%$), with at least $40.0\%$ improvement compared to all the other schemes at 
$\text{SNR}=36\,$dB. 
At $\text{SNR}=38\,$dB, ISAC also achieves $100\%$ success rate, but its trajectory is $5$ meters longer than PISAC. 
The trajectory length, average acceleration, and maximum acceleration of PISAC are the smallest among all simulated schemes, which ensures efficient and smooth driving.

\begin{figure}[!t]
\centering 
\includegraphics[width=0.95\linewidth]{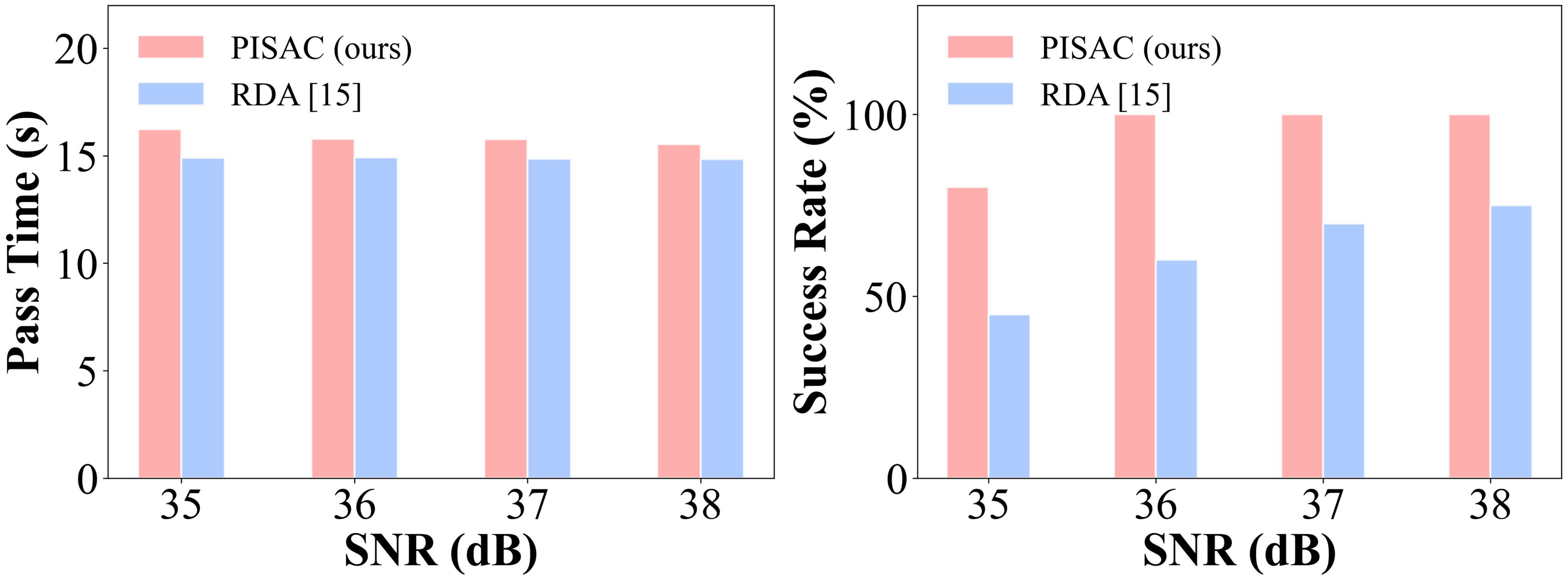}
\caption{Pass times and success rates of PISAC and RDA.}
\vspace{-0.15in}
\label{RDA_PISAC}
\end{figure}

We also compare PISAC with RDA in Fig.~\ref{RDA_PISAC}. 
Since RDA ignores sensing uncertainties, it leads to more aggressive driving and shorter pass time. 
However, the cost is a severe compromise in security. 
Particularly, the success rate can drop below $50\%$, representing an unacceptable risk for CAV systems. 
This experiment demonstrates the necessity of adopting chance constraints \eqref{chance} for planning.

\begin{figure}[!t]
\centering 
\includegraphics[width=0.95\linewidth]{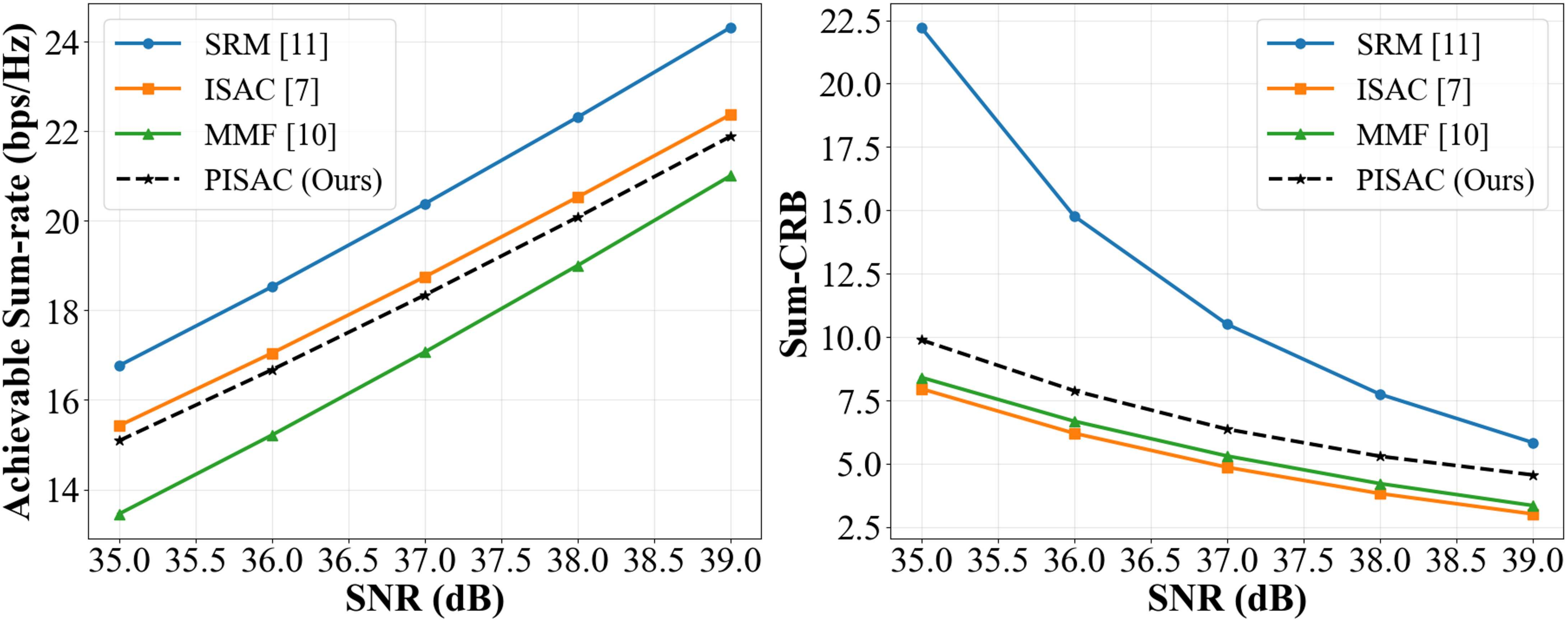}
\caption{Rates and CRBs versus SNR.}
\label{fig4}
\vspace{-0.26in}
\end{figure}

Finally, the sensing and communication performances are also provided in Fig.~\ref{fig4}.
The sum rate of PISAC is $4\%$ to $12\%$ higher than that of the MMF, and the CRB is $22\%$ to $60\%$ lower than that of the SRM. 
This shows balanced and satisfactory performance of PISAC on non-planning metrics.

\vspace{-0.05in}
\section{Conclusion}\label{section6}

This paper presented PISAC for CAVs, which optimizes motion-planning performance under joint sensing and communication constraints. The proposed method employs a safety-bound formulation and a safety-space shrinkage thresholding model to \textbf{identify planning-bottleneck obstacles}. A bilevel PAMP algorithm integrating DCA and ADMM was developed to efficiently solve the coupled optimization. Simulation results demonstrated that PISAC achieves superior trajectory efficiency, safety, and robustness compared with benchmarks.

\vspace{-0.1in}
\bibliographystyle{IEEEtran}
\bibliography{main}

\begin{thebibliography}{10}
\providecommand{\url}[1]{#1}
\csname url@samestyle\endcsname
\providecommand{\newblock}{\relax}
\providecommand{\bibinfo}[2]{#2}
\providecommand{\BIBentrySTDinterwordspacing}{\spaceskip=0pt\relax}
\providecommand{\BIBentryALTinterwordstretchfactor}{4}
\providecommand{\BIBentryALTinterwordspacing}{\spaceskip=\fontdimen2\font plus
\BIBentryALTinterwordstretchfactor\fontdimen3\font minus \fontdimen4\font\relax}
\providecommand{\BIBforeignlanguage}[2]{{%
\expandafter\ifx\csname l@#1\endcsname\relax
\typeout{** WARNING: IEEEtran.bst: No hyphenation pattern has been}%
\typeout{** loaded for the language `#1'. Using the pattern for}%
\typeout{** the default language instead.}%
\else
\language=\csname l@#1\endcsname
\fi
#2}}
\providecommand{\BIBdecl}{\relax}
\BIBdecl

\bibitem{Neupan}
R.~Han \emph{et~al.}, ``{NeuPAN}: Direct point robot navigation with end-to-end model-based learning,'' \emph{IEEE Trans. Robot.}, vol.~41, pp. 2804--2824, Mar. 2025.

\bibitem{9561612}
Z.~Zhang \emph{et~al.}, ``Distributed dynamic map fusion via federated learning for intelligent networked vehicles,'' in \emph{Proc. ICRA}, Xi'an, China, May. 2021, pp. 953--959.

\bibitem{pei2023collaborative}
L.~Pei, J.~Lin, Z.~Han, L.~Quan, Y.~Cao, C.~Xu, and F.~Gao, ``Collaborative planning for catching and transporting objects in unstructured environments,'' \emph{IEEE Robot. Autom. Lett.}, vol.~9, no.~2, pp. 1098--1105, Feb. 2024.

\bibitem{EdgeV2X}
Z.~Li \emph{et~al.}, ``Edge-assisted {V2X} motion planning and power control under channel uncertainty,'' \emph{IEEE Trans. Veh. Technol.}, vol.~72, no.~7, pp. 9641--9646, Jul. 2023.

\bibitem{9793623}
B.~Wang and R.~Su, ``A distributed platoon control framework for connected automated vehicles in an urban traffic network,'' \emph{IEEE Trans. Control Netw. Syst.}, vol.~9, no.~4, pp. 1717--1730, Dec. 2022.

\bibitem{RdarSignalProcess}
F.~Liu, W.~Yuan, C.~Masouros, and J.~Yuan, ``Radar-assisted predictive beamforming for vehicular links: Communication served by sensing,'' \emph{IEEE Trans, Wireless Commun.}, vol.~19, no.~11, pp. 7704--7719, 2020.

\bibitem{Radar_Centric}
F.~Liu and C.~Masouros, ``Joint localization and predictive beamforming in vehicular networks: Power allocation beyond water-filling,'' in \emph{Proc. ICASSP}, Toronto, Canada, Jun. 2021, pp. 8393--8397.

\bibitem{RadarEstimate}
------, ``A tutorial on joint radar and communication transmission for vehicular networks—{Part III}: Predictive beamforming without state models,'' \emph{IEEE Commun. Lett.}, vol.~25, no.~2, pp. 332--336, Feb. 2021.

\bibitem{Meng2023VehicularConnectivity}
X.~Meng, F.~Liu, C.~Masouros, W.~Yuan, Q.~Zhang, and Z.~Feng, ``Vehicular connectivity on complex trajectories: Roadway-geometry aware isac beam-tracking,'' \emph{{IEEE Trans. Wireless Commun.}}, vol.~22, no.~11, pp. 7408--7423, Nov. 2023.

\bibitem{Dou2024IntegratedSensing}
C.~Dou, N.~Huang, Y.~Wu, L.~Qian, Z.~Shi, and T.~Q. Quek, ``Integrated sensing and communication enabled multidevice multitarget cooperative sensing: A fairness-aware design,'' \emph{IEEE Internet Things J.}, vol.~11, no.~17, pp. 29\,190--29\,201, Sep. 2024.

\bibitem{zhang2024efficient}
Q.~Zhang, M.~Shao, T.~Zhang, G.~Chen, J.~Liu, and P.~Ching, ``{An efficient sum-rate maximization algorithm for fluid antenna-assisted ISAC system},'' \emph{IEEE Commun. Lett.}, vol.~29, no.~1, pp. 200--204, 2024.

\bibitem{schulman2014motion}
J.~Schulman, Y.~Duan, J.~Ho, A.~Lee, I.~Awwal, H.~Bradlow, J.~Pan, S.~Patil, K.~Goldberg, and P.~Abbeel, ``Motion planning with sequential convex optimization and convex collision checking,'' \emph{Int. J. Rob. Res.}, vol.~33, no.~9, pp. 1251--1270, Jun. 2014.

\bibitem{TwoStagePlanner}
F.~Eiras, M.~Hawasly, S.~V. Albrecht, and S.~Ramamoorthy, ``A two-stage optimization-based motion planner for safe urban driving,'' \emph{IEEE Trans. Robot.}, vol.~38, no.~2, pp. 822--834, Apr. 2021.

\bibitem{CVXPY}
S.~Diamond and S.~Boyd, ``{CVXPY}: A python-embedded modeling language for convex optimization,'' \emph{J. Mach. Learn. Res.}, vol.~17, no.~83, pp. 1--5, Apr. 2016.

\bibitem{RDA}
R.~Han \emph{et~al.}, ``{RDA}: An accelerated collision free motion planner for autonomous navigation in cluttered environments,'' \emph{IEEE Robot. Autom. Lett.}, vol.~8, no.~3, pp. 1715--1722, Mar. 2023.

\bibitem{dosovitskiy2017carla}
A.~Dosovitskiy \emph{et~al.}, ``{CARLA}: An open urban driving simulator,'' in \emph{Proc. CoRL}, California, USA, Nov. 2017, pp. 1--16.

\bibitem{Tensor_mMIMO}
R.~Zhang, L.~Cheng, S.~Wang, Y.~Lou, Y.~Gao, W.~Wu, and D.~W.~K. Ng, ``Integrated sensing and communication with massive mimo: A unified tensor approach for channel and target parameter estimation,'' \emph{IEEE Trans. Wireless Commun.}, vol.~23, no.~8, pp. 8571--8587, Aug. 2024.

\bibitem{ChanceConstraint}
Y.~K. Nakka and S.-J. Chung, ``Trajectory optimization of chance-constrained nonlinear stochastic systems for motion planning under uncertainty,'' \emph{IEEE Trans. Robot.}, vol.~39, no.~1, pp. 203--222, Feb. 2023.

\bibitem{ChainRule}
Z.~Zhang \emph{et~al.}, ``Multiple intelligent reflecting surfaces collaborative wireless localization system,'' \emph{IEEE Trans. Wireless Commun.}, vol.~24, no.~1, pp. 134--148, Jan. 2025.

\bibitem{MinSum_1}
I.-K. Lee, M.-S. Kim, and G.~Elber, ``Polynomial/rational approximation of minkowski sum boundary curves,'' \emph{Graph. Model. Im. Process.}, vol.~60, no.~2, pp. 136--165, Mar. 1998.

\bibitem{li2024edge}
G.~Li \emph{et~al.}, ``Edge accelerated robot navigation with collaborative motion planning,'' \emph{IEEE/ASME Trans. Mechatron.}, vol.~30, no.~2, pp. 1166--1178, Apr. 2025.

\bibitem{OBCA}
X.~Zhang, A.~Liniger, and F.~Borrelli, ``Optimization-based collision avoidance,'' \emph{IEEE Trans. Control Syst. Technol.}, vol.~29, no.~3, pp. 972--983, May. 2021.

\end{thebibliography}

\end{document}